\documentclass[a4paper]{JHEP3}

\usepackage[latin1]{inputenc}
\usepackage{bbm}
\usepackage{bm}
\usepackage{graphicx}
\usepackage{epsfig}
\usepackage{subfigure}
\usepackage{latexsym}
\usepackage{amsmath}
\usepackage{amsfonts}
\usepackage{amssymb}

\newcommand{\Eslash}{\ensuremath \raisebox{0.025cm}{\slash}\hspace{-0.28cm} E}
\newcommand{\be}{\begin{eqnarray}} 
\newcommand{\ee}{\end{eqnarray}}

\newcommand{\bmp}{\noindent\begin{minipage}{16cm}}
\newcommand{\emp}{\end{minipage}\vskip 7mm} 
\def\lsim{\mathrel{\raise.3ex\hbox{$<$\kern-.75em\lower1ex\hbox{$\sim$}}}}
\def\gsim{\mathrel{\raise.3ex\hbox{$>$\kern-.75em\lower1ex\hbox{$\sim$}}}}

\newcommand{\intron}[1]{}

\title{Natural fourth generation of leptons}

\author{Oleg Antipin\footnote{oleg.a.antipin@jyu.fi}, Matti Heikinheimo\footnote{matti.p.s.heikinheimo@jyu.fi} and Kimmo Tuominen\footnote{kimmo.tuominen@phys.jyu.fi}\\
Department of Physics, University of Jyv\"askyl\"a, Finland and Helsinki Institute of Physics, University of Helsinki, Finland}

\abstract{We consider implications of a fourth generation of leptons, allowing for the most general mass patterns for the fourth generation neutrino. We determine the constraints due to the precision electroweak measurements and outline the signatures to search for at the LHC experiments. As a concrete framework to apply these results we consider the minimal walking technicolor (MWTC) model where the matter content, regarding the electroweak quantum numbers, corresponds to a fourth generation.}


\begin{document}


\section{Introduction and summary}

The experimental program at the Large Hadron Collider (LHC) at CERN may lead to discovery of new physics beyond the Standard Model (SM) of elementary particle interactions soon. Some hints about the insufficience of the Standard model have been already provided by e.g. observation of neutrino masses and, more recently, by the measurements concerning heavy flavors \cite{Lunghi:2009sm,Soni:2008bc}. More definite conclusions are expected after the Higgs sector of SM becomes subject to intensive probing at the LHC. Decades of intensive research on beyond Standard model (BSM) model building and phenomenology have provided several alternative theoretical concepts for experimental testing. Possible model building paradigms include extra dimensions, supersymmetry, unified models and technicolor. 

One of the simplest extensions of the standard model, arguably, is the hypothetical existence of a fourth generation of elementary matter fermions \cite{Frampton:1999xi,Holdom:2006mr,Bobrowski:2009ng}. Since we do not know about the origins of flavor, there seems to be no need to stick to just three generations. However, the {\em{ad hoc}} insertion of yet another replica of quark and lepton doublets and singlets may seem, admittedly, a very unimaginative direction to proceed. The phenomenological appeal for just three generations is due to the close link between the constraints on the number of neutrino species from the formation of light elements in the early universe, i.e. the Big Bang nucleosynthesis and the high-energy experiments measuring the $Z$-width. Taken these together, seems to show indisputably that there are only three conventional neutrinos with mass below $M_Z/2\simeq 45$ GeV. However, the need for BSM physics provides  phenomenological tension towards the other direction and this has led to a large variety of further model building resting on the introduction of new degrees of freedom. For example, in grand-unified theories (GUTs) beyond the minimal SU(5) case there inevitably appear new fermions and models with supersymmetry or new gauge interactions introduce, by definition, extended particle spectrum. 

In this paper, as a general framework, we consider (walking) technicolor \cite{TC,Hill:2002ap,Holdom:1981rm,Yamawaki:1985zg,Appelquist:an}, and as a particular realization we take the minimal walking technicolor (MWTC) model \cite{Sannino:2004qp}. In this model the electroweak symmetry breaking is due to strong dynamics of two Dirac fermions (techniquarks) transforming in the adjoint representation of SU(2) technicolor gauge group. This particular matter content is minimal in the sense that with relatively small number of new matter fields, this model has been proposed to lie close to an infrared fixed point \cite{Sannino:2004qp} which renders the coupling almost conformal over large hierarchy of scales. Phenomenologically such feature is welcome since on one hand walking is required to tame the effects of flavor changing neutral currents and on the other hand the minimal matter content is necessary to keep the contributions to the precision $S$-parameter on the level compatible with observations \cite{Dietrich:2005jn}. To strengthen the phenomenological viability of this model, it has recently been studied from first principles on the lattice by measuring the properties of the physical spectrum \cite{Catterall:2007yx, Catterall:2008qk, Hietanen:2008mr} and the evolution of the coupling constant \cite{Hietanen:2009az}. These studies point to the conclusion that this theory is indeed near conformal as originally proposed in \cite{Sannino:2004qp}, and hence provides a good model building basis for walking Technicolor -type theory.

The matter field spectrum of the MWTC model, from the electroweak interaction viewpoint, features three (techni)quark doublets. Due to a global anomaly \cite{Witten:1982fp}, such particle content results in an ill defined theory, but this anomaly can be simply cured by introducing just one further doublet taken to be singlet under QCD and technicolor interactions and this doublet therefore resembles the leptons of ordinary three SM generations. Hence, by the internal consistency of the underlying gauge theory, we are led to consider a model with a somewhat nonstandard fourth generation. Several phenomenological constraints on the properties of this fourth generation exist. Since the interactions of the techniquarks become strong at the scale of the order of $v\simeq$250 GeV, they are confined inside technihadrons which the past and present colliders simply were not able to produce. With the leptons on one hand, we know that the analogue of the ordinary electron has to be very massive, few times $M_Z$ at least. For the fourth generation neutrino the issue is more subtle: For example, consider the fourth neutrino as a Dirac particle of mass $\sim M_Z/2$. Then, if it is unstable it is ruled out by the LEP II bounds and if it is stable, then it contributes to the dark matter abundance and is ruled out by the CDMS experiment \cite{Akerib:2004fq,Kainulainen:2006wq}. For a stable purely left-handed Majorana neutrino the constraints are weaker due to smaller cross sections. However, the dark matter searches have ruled out a purely left-handed sequential fourth generation Majorana neutrino up to masses of the order of a few TeV \cite{Angle:2008we}. These bounds are alleviated if more general mixing patterns in the neutrino sector are allowed for \cite{Kouvaris:2007iq,KTVII}. In particular, if the lightest state is dominated by the weak singlet component, then its couplings to standard model are further weakened, allowing it to escape detection so far even for relatively small masses below ${\mathcal{O}}(M_Z/2)$. Such mix
ing patterns will be important in relation to the dark matter a bundance \cite{KTVII}; here we concentrate on the the collider phenomenology aspects of this model and do not require absolute stability of the fourth generation neutrino.

Various aspects of the phenomenology implied by this model have already been investigated in the literature. For the technihadronic sector, see e.g. \cite{Foadi:2007ue}. The leptonic sector is particularly interesting since its contributions to the precision observables can be perturbatively evaluated, and existing data can be used to constrain the masses of these leptons. This analysis has been carried out for the cases in which the fourth generation neutrino has only a Dirac mass \cite{Dietrich:2005jn,He:2001tp}, is a purely left-handed Majorana state \cite{Holdom:1996bn,Kainulainen:2006wq} or has a particular mixing pattern between left- and right-handed neutrinos \cite{Bertolini:1990ek,Gates:1991uu,Kniehl:1992ez}. In this paper we extend these studies to allow for the most general mass and mixing patterns of the fourth generation neutrino. Our results are general and provide, to our knowledge, a so far unexplored completion of the existing literature.

Furthermore, we apply this analysis to the MWTC model in order to identify the phenomenologically most interesting mass ranges of the fourth generation leptons. Given these, we investigate several possible signals which should be of interest at the LHC. In particular we emphasize important differences between the minimal technicolor model and models where a sequential full fourth SM-like generation is considered, see e.g. \cite{Kribs:2007nz}: For example, there is no fourth QCD-quark generation and therefore the Higgs production through gluon fusion is not enhanced in the case of technicolor. However, the Higgs can decay into fourth generation neutrino, which has to be massive to avoid observation so far, and creates a new channel which will, for neutrino masses $\sim {\mathcal{O}}(M_z/2)$, diminish other channels expected to be relevant for light Higgs on the basis of the Standard Model or the onset of the channels with $ZZ$ and $WW$ final states if the new neutrino has mass of the order of $M_Z$. Together with these results, the fourth generation leptons with masses in the range accessible at LHC provide clear direct signals already at first 10 fb$^-1$ of integrated luminosity as our analysis shows.     

The paper is organized as follows: In sec. \ref{model} we first present the details of the minimal walking technicolor, in particular its leptonic sector. In sec. \ref{results} we first present a general analysis of the oblique corrections for a lepton generation with massive neutrino and apply it to the MWTC model. Then we discuss collider signatures for the production of new leptons and how they may affect the Higgs production and decay rates. We conclude and outline some future directions in sec. \ref{checkout}.

\section{The model}
\label{model}
As discussed in the introduction, the present Technicolor model building rests on two paradigms: On one hand, walking of the coupling is required in order to suppress the flavor changing neutral current interactions which will arise if the Technicolor model is embedded into some extended Technicolor framework. On the other hand, the walking must be achieved with reasonably small number, say  two or three, of techniquark flavors in order to not generate too large contributions to the precision observable $S$. It has been proposed \cite{Sannino:2004qp} that $SU(2)$ gauge theory with two fermion flavors in the adjoint representation of the gauge group is a minimal candidate for such a theory \footnote{For an ultraminimal alternative, see \cite{Ryttov:2008xe}}. Let us now build up the concrete model Lagrangian, starting with the Technicolored sector
\begin{eqnarray}
{\mathcal{L}}_{\rm{TC}}=-\frac{1}{4}{\mathcal{F}}^a_{\mu\nu}{\mathcal{F}}^{a\mu\nu}
+i\overline{Q}_L\gamma^\mu D_\mu{Q}_L+i\overline{U}_R\gamma^\mu D_\mu{U}_R
+i\overline{D}_R\gamma^\mu D_\mu{D}_R,
\label{TC_lagrangian}
\end{eqnarray}
where ${\mathcal{F}}^a_{\mu\nu}=\partial_\mu{\mathcal{A}}^a_\nu-\partial_\nu{\mathcal{A}}^a_\mu
+ig_{\rm{TC}}\epsilon^{abc}{\mathcal{A}}^b_\mu{\mathcal{A}}^c_\nu$ is the usual field strength, $a=1,2,3$, and the $SU_L(2)$ weak doublet $Q^T_L=(U_L,D_L)$ while $U_R$ and $D_R$ are singlets under the weak isospin. 
The number of weak doublets in this theory is odd, rendering the gauge theory $SU_L(2)$ of weak interactions anomalous. To cure this anomaly at least one weak doublet is needed. Since the walking behavior of the Technicolor theory should not be spoiled, we add a doublet uncharged under technicolor $SU(2)$ gauge group. To add just one doublet, we cannot assign QCD color either, and hence this doublet resembles a new generation of ordinary leptons. The anomaly free hypercharge assignments are 
\begin{eqnarray}
Y(Q_L) &=& y/2, ~~Y(U_R,D_R)=((y+1)/2,(y-1)/2) \nonumber \\
Y(L_L) &=& -3y/2, ~~Y(\zeta_R, \nu_{\zeta,R})=((-3y+1)/2,(-3y-1)/2),\nonumber
\end{eqnarray}
where $y$ is any real number. A particular choice $y=1/3$ corresponds to a standard model -like fourth family. Other choices are possible, but here we confine ourselves to $y=1/3$, and we will be mostly interested in the phenomenology arising from the leptonic sector.
\begin{eqnarray}
{\mathcal{L}}_\ell=i\overline{L}_L\gamma^\mu D_\mu L_L+i\overline{E}_R\gamma^\mu D_\mu E_R
+i\overline{N}_R\gamma^\mu D_\mu N_R.
\label{lepton_lagrangian}
\end{eqnarray}
However, instead of the Lagrangian (\ref{TC_lagrangian}), at the electroweak scale the Technicolor part is better described by the chiral effective theory coupled to the electroweak gauge fields and matter fields. Such chiral effective theory can also be systematically improved, and also additional degrees of freedom like the vectors and axial vectors can be included \cite{Foadi:2007ue}. Technicolor theories are constructed to describe only the mass patterns of the electroweak gauge bosons, and one needs to address the question of the masses of the elementary matter fields separately. One traditional direction which has been pursued in the literature is so called extended technicolor (ETC), which couples technifermions and ordinary SM fermions with each other by extended gauge interactions assumed broken at some high scale $M_{\textrm{ETC}}\gg \Lambda_{\textrm{TC}}$ and described at energies below $M_{\textrm{ETC}}$ by effective four-fermion interactions. Condensation of technifermions then leads to mass terms for the SM fermions. Since we do not know the ultraviolet complete gauge theory possibly underlying fermion mass generation as described above, we choose here a more modest bottom-up approach in order to parametrize our ignorance of the origin of mass for the matter fields in terms of effective Yukawa interactions between the fermions and the Higgs. In MWTC model, with the hypercharge assignments we are using here, the hypercharge conservation allows coupling only between the SM matter fields and the technimeson with quantum numbers of the SM-like Higgs scalar.   
Hence, to estimate the effects of the scalar sector on the new leptons up to and including dimension five operators, we consider following effective interactions \cite{Foadi:2007ue}.
\begin{eqnarray}
& &{\mathcal{L}}_{{\rm{Mass}}} = (y \bar{L}_L H E_R+ {\rm{h.c.}})+C_D\bar{L}_L\tilde{H}N_{R}\nonumber \\
&+& \frac{C_{L}}{\Lambda}(\bar{L}^c\tilde{H})(\tilde{H}^TL)+\frac{C_{R}}{\Lambda}(H^\dagger H)\bar{N}^c_{R}N_{R}+{\rm{h.c.}} 
\label{scalar_fermion}
\end{eqnarray}
where $\tilde{H}=i\tau^2H^\ast$. The first term in (\ref{scalar_fermion}) lead to the usual (Dirac) mass for the charged fourth generation lepton, and the remaining terms allow for the most general mass structure of the fourth neutrino.   

After symmetry breaking the resulting neutrino mass terms are
\begin{eqnarray}  
-\frac{1}{2}\bar{n}_L^{~c}Mn_L+{\rm{h.c.}},~~~M=\left(\begin{array}{cc} M_L & m_D \\ m_D & M_R\end{array}\right),
\end{eqnarray}
where $n_L=(N_{L}, N_{R}^{~c})^T$, $m_D=C_Dv/\sqrt{2}$ and $M_{L,R}=C_{{L,R}}v^2/2\Lambda$. The scale $\Lambda$ is of the order of 1 TeV.
The special cases are pure Dirac and pure left-handed Majorana neutrino which are obtained, respectively, by discarding dimension five operators and by removing the right handed field $N_{R}$. In the general case there are two Majorana eigenstates, $\chi_1$ and $\chi_2$ associated with the eigenvalues
\begin{eqnarray}
\lambda_{1,2}=\frac{1}{2}\left[(M_L+M_R)\pm\sqrt{(M_L-M_R)^2+4m_D^2}\right]
\end{eqnarray}
of the mass matrix. Since $\lambda$ may be positive or negative, we define $\lambda_k=M^\prime_k\equiv M_k\rho_k$, where $\rho_k=\pm 1$ so that $M_k>0$ is ensured. Note that there are basically two equivalent ways to treat the $\rho$-factors \cite{Barroso:1983rd}. Here we will choose to include these factors into the definition of the transformation into the mass eigenbasis. The advantage of this approach is that the $\rho$-dependence will show on the Lagrangian level explicitly. Another alternative is to include the $\rho$-factors into the definition of the Majorana field operators and then one must keep track of the appearance of these factors when evaluating the contractions corresponding to individual Feynman diagrams. In order to maintain full generality, we keep track of these phases explicitly and present the results for the charged and neutral weak currents as well as for the couplings to the composite Higgs in detail below. The following discussion has been adapted from \cite{KTVII} where a similar derivation was carried out for the first time.

The mass eigenstates are obtained with the diagonalizing matrix, 
\begin{equation}
{\mathcal{U}}=\left(\begin{array}{cc} \sqrt{\rho_1}\cos\theta & \sqrt{\rho_2}\sin\theta \\
-\sqrt{\rho_1}\sin\theta & \sqrt{\rho_2}\cos\theta \end{array}\right),
\end{equation}
and the eigenstates are 
\begin{eqnarray}
\chi={\mathcal{U}}^\dagger n_L+{\mathcal{U}}^Tn_L^c,
\end{eqnarray}
Note that with this prescription $\chi_k^c=\chi_k$, since $\chi_k$ is a Majorana state with mass $M_k$ by construction. 
The mixing angle $\theta$ is given by $\tan(2\theta)=2m_D/(M_R-M_L)$.

In the mass eigenbasis the gauge interactions are
\begin{eqnarray}
W^+_\mu\bar{N}_L\gamma^\mu E_L &=& \frac{\cos\theta}{\sqrt{\rho_1}}\bar{\chi}_{1L}W^+_\mu\gamma^\mu E_L +\frac{\sin\theta}{\sqrt{\rho_2}}\bar{\chi}_{2L}W^+_\mu\gamma^\mu E_L\nonumber \\
Z_\mu\bar{N}_L\gamma^\mu N_L &=& \cos^2\theta Z_\mu\bar{\chi}_{1L}\gamma^\mu \chi_{1L}+\sin^2\theta Z_\mu\bar{\chi}_{2L}\gamma^\mu \chi_{2L}\nonumber \\
&& +\frac{1}{2}\sin(2\theta)Z_\mu(\frac{1}{\sqrt{\rho_2}^\ast\sqrt{\rho_1}}\bar{\chi}_{1L}\gamma^\mu \chi_{2L}+\frac{1}{\sqrt{\rho_2}\sqrt{\rho_1}^\ast}\bar{\chi}_{2L}\gamma^\mu \chi_{1L}) 
\label{gauge_interactions}
\end{eqnarray}
The last terms in the neutral current can be combined into
\begin{eqnarray}
\frac{1}{\sqrt{\rho_2}\sqrt{\rho_1}^\ast}(\bar{\chi}_2\gamma^\mu P_L\chi_1+(\sqrt{\rho_2}\sqrt{\rho_1}^\ast)^2\bar{\chi}_1\gamma^\mu P_L\chi_2)=\frac{1}{\sqrt{\rho_1}^\ast\sqrt{\rho_2}}\bar{\chi}_2\gamma^\mu(\alpha-\beta\gamma_5)\chi_1,
\end{eqnarray}
where $\alpha=\frac{1}{2}(1-(\sqrt{\rho_1}^\ast\sqrt{\rho_2})^2)$ and $\beta=\frac{1}{2}(1+(\sqrt{\rho_1}^\ast\sqrt{\rho_2})^2)$. 

The effective interactions between the Higgs and neutrino following from (\ref{scalar_fermion}) are
\begin{eqnarray}
\frac{h}{2v}\bar{n}_L^c\left(\begin{array}{cc} 2M_L & m_D \\ m_D & 2M_R\end{array}\right)n_L 
+{\rm{h.c.}},
\label{nu_higgs_L}
\end{eqnarray}
where interaction terms of ${\mathcal{O}}(h^2)$ have been neglected, since we do not need them in the following because we will be interested in the vertices relevant for the decay of the Higgs to the new neutrinos. Translating into the mass eigenbasis we obtain
\begin{eqnarray}
{\mathcal{L}}_{\rm{Higgs}}=C_{22}h\bar{\chi}_2\chi_2+C_{11}h\bar{\chi}_1\chi_1+C_{21}h\bar{\chi}_1(\beta+\alpha\gamma_5)\chi_2+\dots
\label{higgs_interactions}
\end{eqnarray}
where we have defined
\begin{eqnarray}
C_{11} &=&  \frac{M_1}{v}(1-\frac{1}{4}\sin^2(2\theta)(1-(\sqrt{\rho_1}^\ast\sqrt{\rho_2})^2\frac{M_2}{M_1})), \nonumber \\
C_{22} &=&  \frac{M_2}{v}(1-\frac{1}{4}\sin^2(2\theta)(1-(\sqrt{\rho_1}\sqrt{\rho_2}^\ast)^2\frac{M_1}{M_2})), \nonumber \\
C_{12} &=& -\frac{M_2}{4v}\sqrt{\rho_1}\sqrt{\rho_2}^\ast\sin(4\theta)(1-(\sqrt{\rho_1}^\ast\sqrt{\rho_2})^2\frac{M_1}{M_2}),
\end{eqnarray}
and the factors $\alpha$ and $\beta$ are the same factors as defined few lines earlier for the neutral current. 

\begin{center}
\begin{figure}
\centerline{\includegraphics[width=10cm]{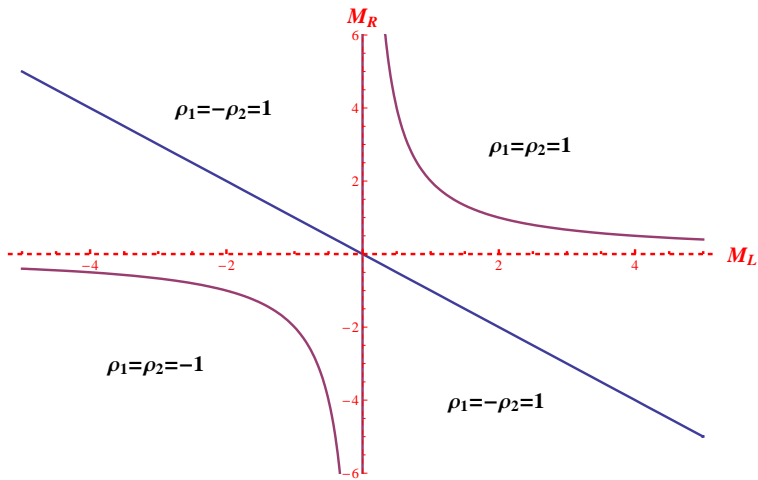}}
\caption{Schematic figure of the parameter space corresponding to a fixed value of $m_D$. The hyperbolas and the straight diagonal line correspond to sections of surfaces $m_D^2=M_LM_R$ and $M_R=-M_L$, respectively. The $m_D$-axis is perpendicular to the $(M_L,M_R)$-plane.}
\label{rho_fig}
\end{figure}
\end{center}

Since the parameters $M_L$, $M_R$ and $m_D$ are simply coupling constants in our formulation, there is no need to restrict to positive values. If we assume that these parameters are real numbers free to take any value, then in terms of the mass eigenvalues and -states as defined above the parameter space contains three domains corresponding to $\rho_1=\rho_2=\pm 1$ and $\rho_1=-\rho_2=1$, and in each case $M_1$ and $M_2$ assume all positive real values and $0\le \sin\theta\le 0.5$ \cite{KTVII}. In Figure \ref{rho_fig} a slice of the parameter space $(M_L,M_R,m_D)$ is shown for some fixed finite value of $m_D$. The hyperbolas correspond to surfaces $m_D^2=M_LM_R$, and together with the plane $M_R=-M_L$ they divide the parameter space into three distinct parts in which the values of the $\rho$-parameters are as indicated in the figure. The parameter space is symmetric with respect to the plane $M_R=-M_L$ with replacement $M_1\leftrightarrow M_2$. It is therefore sufficient to restrict to the upper half corresponding to $M_R\ge -M_L$, and this is also reflected by the fact that the interactions practically only depend on the product of $\rho_1\rho_2$ and not separately on $\rho_1$ and $\rho_2$.  

Typical special cases correspond to setting either $M_L$ or $M_R$ equal to zero, and these both correspond to $\rho_1=-\rho_2=1$. Both $M_L$ and $M_R$ need to be nonzero in order to obtain same sign for the $\rho$-parameters. Our results for the currents in (\ref{gauge_interactions}) as well as the Higgs interactions in (\ref{nu_higgs_L}) can be applied for any value of $M_L$, $M_R$ and $m_D$ with appropriate choice of $\rho$-parameters. For the mixing phenomena we stress the following general feature: using the above formulae it can be shown that in the case $M_L=0$ the lighter state is always dominated by the weak doublet component $\nu_L$ and hence couples stronger to the electroweak currents than the heavy state. On the other hand, for $M_R=0$ the lighter state consists dominantly of the weak singlet component and in this case the lighter component has weaker coupling to the electroweak currents. A general feature therefore is that a large hierarchy $M_L\gg M_R$ will make the lighter neutrino state more difficult to observe since its couplings to the weak currents are diminished. In this part of the parameter space the heavier neutrino state might be phenomenologically more accessible provided that its mass is still within reach at the LHC. And vice versa, in the parameter space domain corresponding to $M_L\ll M_R$ the lighter neutrino state should be better accessible than the heavy one. Let us then turn to the phenomenological implications. 

\section{Results}
\label{results}
In this section we will present the phenomenological results of the model. We first study the oblique corrections $S$ and $T$. This study could be enlarged to include full set of electroweak precision parameters \cite{Barbieri:2004qk}, but here we concentrate only on these two mainly since the $S$-parameter is known to provide most stringent constraints for the Technicolor models in general. We have also checked that the precision parameter $U$ is small over the parameter range which we consider. Since the data \cite{:2005ema,Amsler:2008zzb} on $S$ and $T$ shows tendency along the direction $S\sim T$, we apply the generic strategy \cite{Peskin:2001rw} of compensating for positive contribution to $S$ from the techniquarks by a positive contribution to $T$ from the mass splitting within the fourth generation leptons. The leptons typically also provide small negative contribution to the $S$ parameter and this further helps to reconcile the model with the data. The main phenomenological goal of the $S,T$-analysis here is therefore to obtain constraints for the mass splittings within the fourth generation leptons. Once we have this information we then proceed to consider plausible collider signatures for the production rates of these leptons and also their possible implications on the Higgs physics.

\subsection{Oblique corrections}

In the literature the oblique corrections have been analyzed for certain special cases of the mass spectrum of the fourth generation neutrinos. In particular, these include the fourth generation neutrino with Dirac mass term \cite{Dietrich:2005jn} ($M_L=M_R=0$), a pure left Majorana state \cite{Holdom:1996bn,Kainulainen:2006wq} ($M_R=m_D=0$) and for the case corresponding to the usual type I seesaw mass matrix \cite{Bertolini:1990ek,Gates:1991uu,Kniehl:1992ez} ($M_L=0$). Here we treat the general case as described in the previous section and present explicitly the formulas required for the evaluation of the oblique corrections. We have checked both analytically and numerically that the results of the above mentioned special cases are properly obtained in the corresponding limits of our formulas. We stress that our results can be applied for any values of $M_L,M_R$ and $m_D$.

Recall the charged and neutral currents from (\ref{gauge_interactions}) involving neutrinos which, together with the usual forms of the currents for the charged lepton allow us to evaluate following contributions to the self energies:
\begin{eqnarray}
\Pi_{3Y}(q^2) &=& -\frac{1}{2}\cos^4\theta\left[\Pi_{LL}(M_1^2,M_1^2,q^2)-\Pi_{LR}(M_1^2,M_1^2,q^2)\right]\nonumber \\
&& -\frac{1}{2}\sin^4\theta\left[\Pi_{LL}(M_2^2,M_2^2,q^2)-\Pi_{LR}(M_2^2,M_2^2,q^2)\right]\nonumber \\
&& 
-\frac{1}{4}\sin^2(2\theta)\left[\Pi_{LL}(M_1^2,M_2^2,q^2)+(-1)^\beta\Pi_{LR}(M_1^2,M_2^2,q^2)\right]\nonumber \\
& & +\Pi_{LR}(M_E^2,M_E^2,q^2)+\frac{1}{2}\Pi_{LL}(M_E^2,M_E^2,q^2),
\end{eqnarray}
which will be needed for the $S$-parameter and
\begin{eqnarray}
\Pi_{11}(0)-\Pi_{33}(0) &=& \frac{1}{2}\cos^2\theta\Pi_{LL}(M_1^2,M_E^2,0)+\frac{1}{2}\sin^2\theta\Pi_{LL}(M_2^2,M_E^2,0)-\frac{1}{4}\Pi_{LL}(M_E^2,M_E^2,0) \nonumber \\
&&-\frac{1}{4}\cos^4\theta\left[\Pi_{LL}(M_1^2,M_1^2,0)-\Pi_{LR}(M_1^2,M_1^2,0)\right] \nonumber \\
&&-\frac{1}{4}\sin^4\theta\left[\Pi_{LL}(M_2^2,M_2^2,0)-\Pi_{LR}(M_2^2,M_2^2,0)\right] \nonumber \\
&&-\frac{1}{8}\sin^2(2\theta)\left[\Pi_{LL}(M_1^2,M_2^2,0)+(-1)^\beta\Pi_{LR}(M_1^2,M_2^2,0)\right] \nonumber \\
\end{eqnarray}
which is needed for the $T$-parameter. The subscripts refer to electroweak gauge boson quantum numbers in the unbroken basis and the relevant Feynman rules for Majorana particles are discussed e.g. in \cite{Denner:1992me}. The vacuum polarizations of the left- and right-handed currents appearing in the above equations are given by
\begin{eqnarray}
\Pi_{LL}(m_1^2,m_2^2,q^2) &=& -\frac{4}{(4\pi)^2}\int_0^1 dx\ln\left[\frac{\mu^2}{M^2-x(1-x)q^2}\right](x(1-x)q^2-\frac{1}{2}M^2)\\
\Pi_{LR}(m_1^2,m_2^2,q^2) &=& -\frac{4}{(4\pi)^2}\int_0^1 dx\ln\left[\frac{\mu^2}{M^2-x(1-x)q^2}\right]\frac{1}{2}m_1m_2,
\end{eqnarray}
where $M^2=xm_1^2+(1-x)m_2^2$. The cutoff $\mu$ has physical significance since we are considering an effective field theory for the generation of Majorana masses of the fourth generation neutrino. 

With these preliminary definitions, the $S$ parameter is given by 
\begin{eqnarray}
-8\pi\frac{d\pi_{3Y}(q^2)}{dq^2}\left|_{q^2=0}\right.\approx -\frac{8\pi}{M_Z^2}(\Pi_{3Y}(M_Z^2)-\Pi_{3Y}(0)),
\end{eqnarray}
and the definition of $T$ is
\begin{eqnarray}
T=\frac{4\pi}{s^2c^2M_Z^2}\left(\Pi_{11}(0)-\Pi_{33}(0)\right),
\end{eqnarray}
where $s^2=\sin^2\theta_W$ and $c^2=\cos^2\theta_W$ with $\theta_W$ the usual weak mixing angle. 

As already mentioned in the previous section it suffices to concentrate only on two regions in the parameter space spanned by the neutrino masses: We assume $M_1>M_2$ and this corresponds to $M_R>-M_L$. This implies that $\rho_1$ is always positive, and the sign of $\rho_2$ is determined by the ratio of $m_D^2$ and $M_RM_L$ so that negative $\rho_2$ corresponds to $m_D^2>M_RM_L$. The sign of $\rho_2$ is reflected in the interaction terms by $\alpha=0$ and $\beta=1$ for $\rho_2=1$ and vice versa for $\rho_2=-1$. Note that the Dirac limit is contained only in the latter domain. In both of these domains all positive values of $M_1$ and $M_2$ as well as all values $0\le \sin\theta\le 1.0$ are allowed, but the difference follows from the property that $\rho_2$ is positive (negative) for $m_D^2>M_LM_R$ ($m_D^2<M_LM_R$). Separately for each of these domains, we explored the parameter space spanned by $M_E,M_1,M_2$ and $\sin\theta$. It is rather obvious that with four parameters it is not difficult to find ranges of values where the experimental constraints from $S$ and $T$ are satisfied. We probed the parameter space in terms of the mass differences of charged and neutral leptons in order to identify the possible spectra consistent with the current experimental bounds. For the technicolor sector we include the naive perturbative estimate $(S,T)=(1/(2\pi),0)\approx (0.16,0)$. 

Replacing the derivative with a finite difference in the definition of the $S$-parameter is a standard approximation known to be valid for new physics with mass scales above $M_Z$. We will consider situations where one neutrino state is lighter than $M_Z$, and one might worry that this is a source for large uncertainty due to the approximation. However this is not the case, since generally the contribution to $S$ from the leptons in the case of one light neutrino is much smaller than from the techniquarks. We have checked that even in the worst case, the error due to the approximation of replacing the derivative with finite difference is at few percent level for the total value of $S$.   
    
\begin{figure}
\centerline{\includegraphics[width=0.7\textwidth]{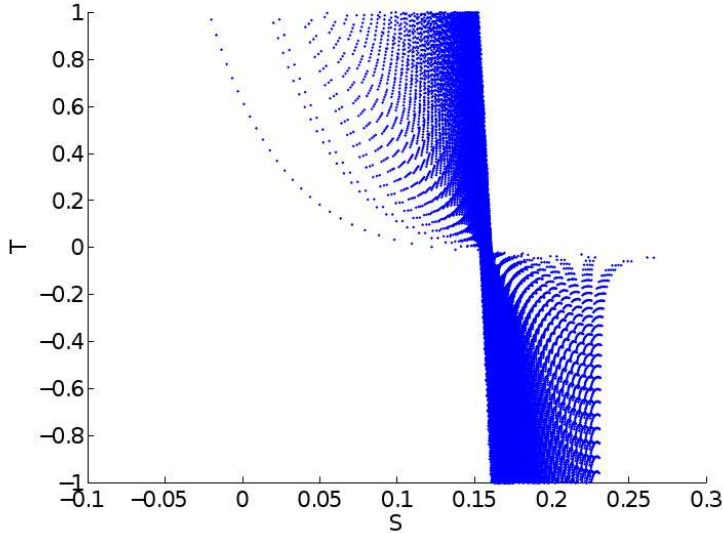}}
\caption{The resulting $S$ and $T$ values as the masses $M_1$, $M_2$ and $M_E$ in the lepton sector vary from 0.5$M_Z$ to 10$M_Z$. The value of the mixing angle is set to $\sin\theta=0.3$.}
\label{ST_scat}
\end{figure}

We found that the results in domains $\rho_2=-1$ and $\rho_2=+1$ are practically identical for the mass ranges of interest. From now on in this section we therefore consider explicitly only the case $\rho_2=-1$. The $S$-parameter is independent of the cutoff as can be directly verified using above definitions. However, for $T$-parameter the scale dependence is more subtle. The divergent contribution can be extracted analytically and it has the simple form $\sim M_L\ln(\mu)$, so in the special cases where $M_L=0$ also $T$ is scale independent. The existence of this divergence signifies the fact that within the model we consider there does not exist a renormalizable Yukawa interaction which would provide the mass for the left-handed Majorana state. We fix the scale by the mass of the heavier neutrino eigenstate, $\mu\propto M_1$, and estimate the uncertainties resulting from the choice of the constant of proportionality. For purely left handed Majorana state this prescription coincides with the one employed in \cite{Holdom:1996bn,Kainulainen:2006wq}. Varying the scale from $\mu=1.5M_1$ to $\mu=2M_1$ results in at most roughly ${\mathcal{O}}(10\%)$ uncertainty in our results concerning the $T$-parameter. As already noted, $S$-parameter does not depend on the scale at all in this case, but it turns out to be far less restrictive than $T$.
With this uncertainty in mind, we fix $\mu=1.5 M_1$ in what follows. In Fig. \ref{ST_scat}, we show the typical scatter plot on the resulting $S$ and $T$ values as the masses $M_1$, $M_2$ and $M_E$ are allowed to vary from 0.5$M_Z$ to 10$M_Z$ with the ordering $M_2<M_1<M_E$, and mixing angle $\sin\theta=0.3$.

\begin{figure}[htb]
\includegraphics[width=0.5\textwidth]{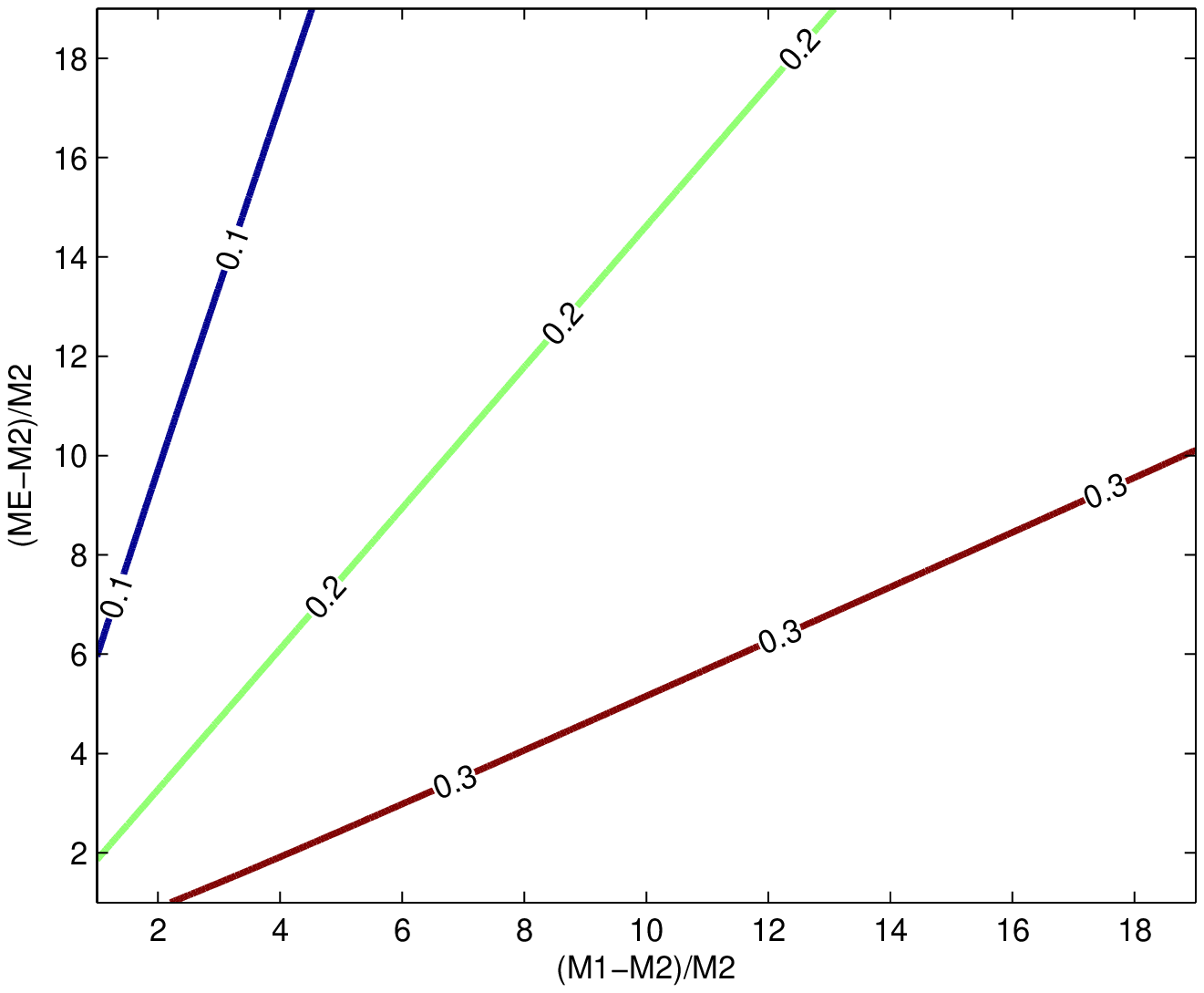}
\includegraphics[width=0.5\textwidth]{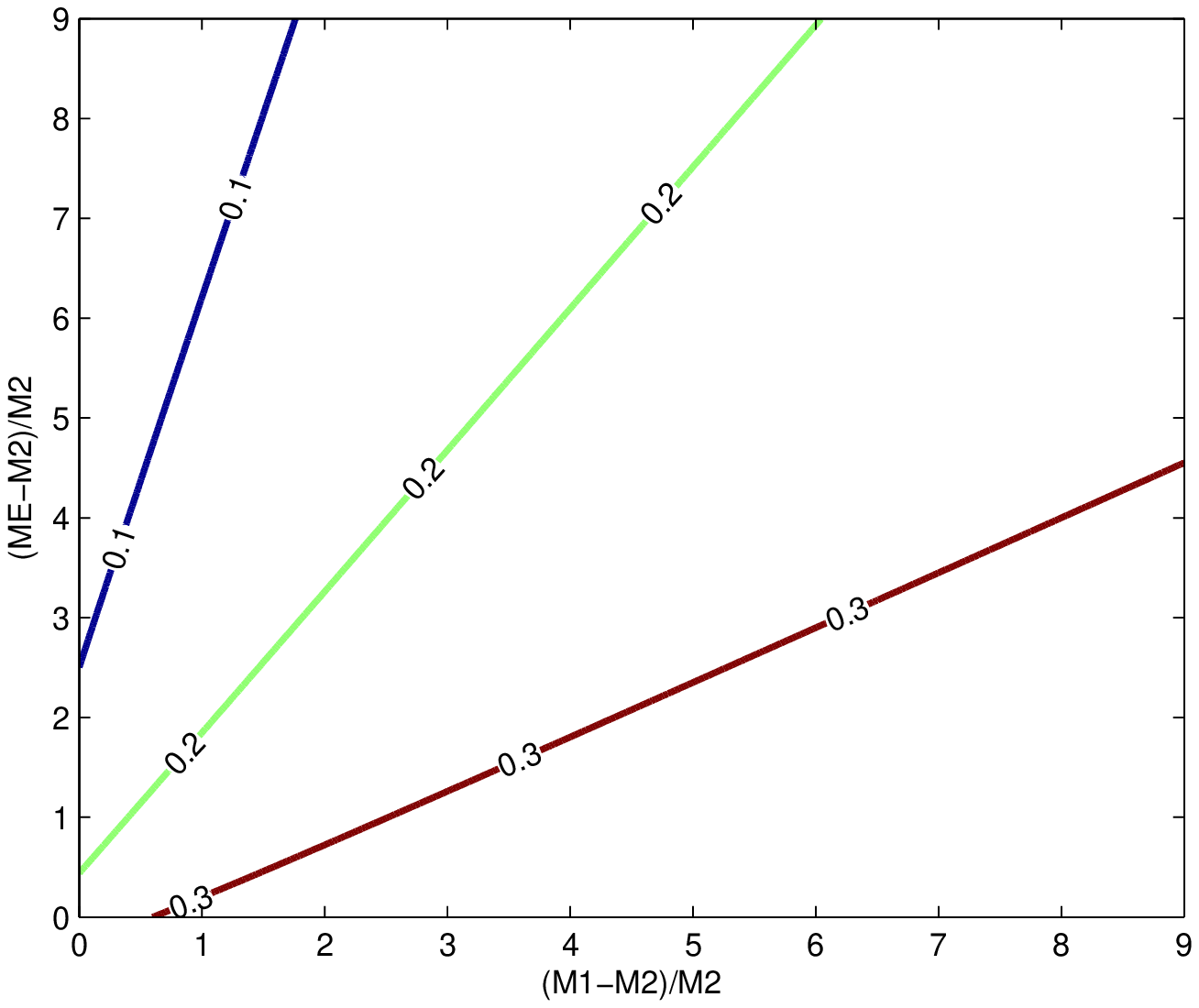}
\caption{Constant $S$ -contours in the $((M_1-M_2),(M_E-M_2)$-plane for the choice of masses $M_2=0.5 M_Z$, (left) and $M_2=M_Z$ (right). The curves correspond to $\sin\theta=0.1$ but the dependence on the angle is very weak.}
\label{Scontours_rhoneg}
\end{figure}

We then investigate the constraints for the parameter space of the model imposed by $S$ and $T$ in more detail taking as guiding limits $|S|\le 0.3$ and $0<T<1$.
We fix the mass of the lighter neutrino state ($M_2$) to be equal to $M_Z/2$ or $M_Z$ and the results for corresponding constant $S$ contours are shown in Fig. \ref{Scontours_rhoneg}. We have checked that the results depend only very weakly on the mixing angle, and hence the curves shown in the figure explicitly for $\sin\theta=0.1$ can be taken as representatives for any value of $\sin\theta$. If we consider larger values of $M_2$, the results lead to similar curves as would be expected since the contributions to both $S$ and $T$ should depend on the relative differences of the masses rather than their absolute values.

\begin{figure}[htb]
\includegraphics[width=0.5\textwidth]{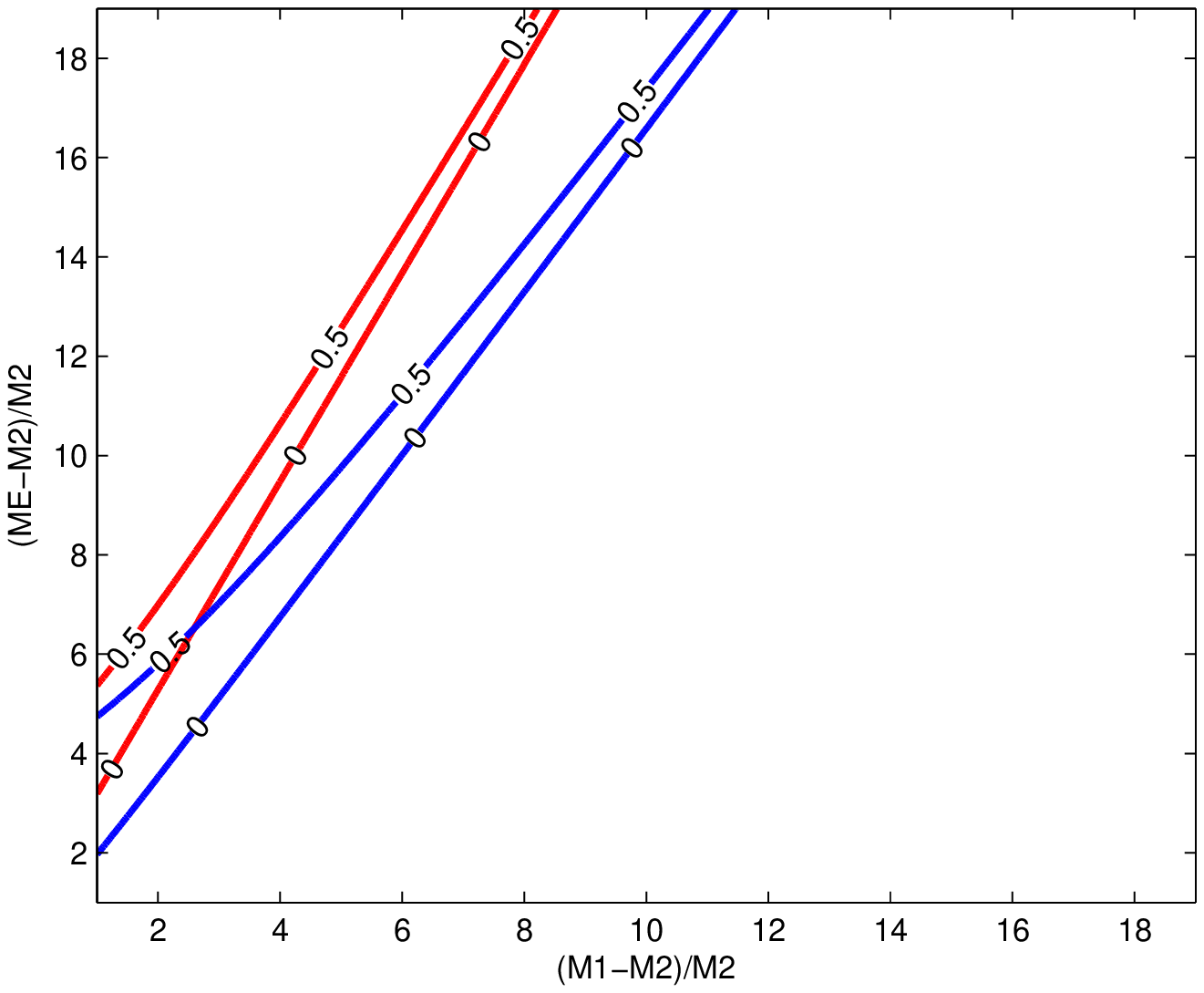}
\includegraphics[width=0.5\textwidth]{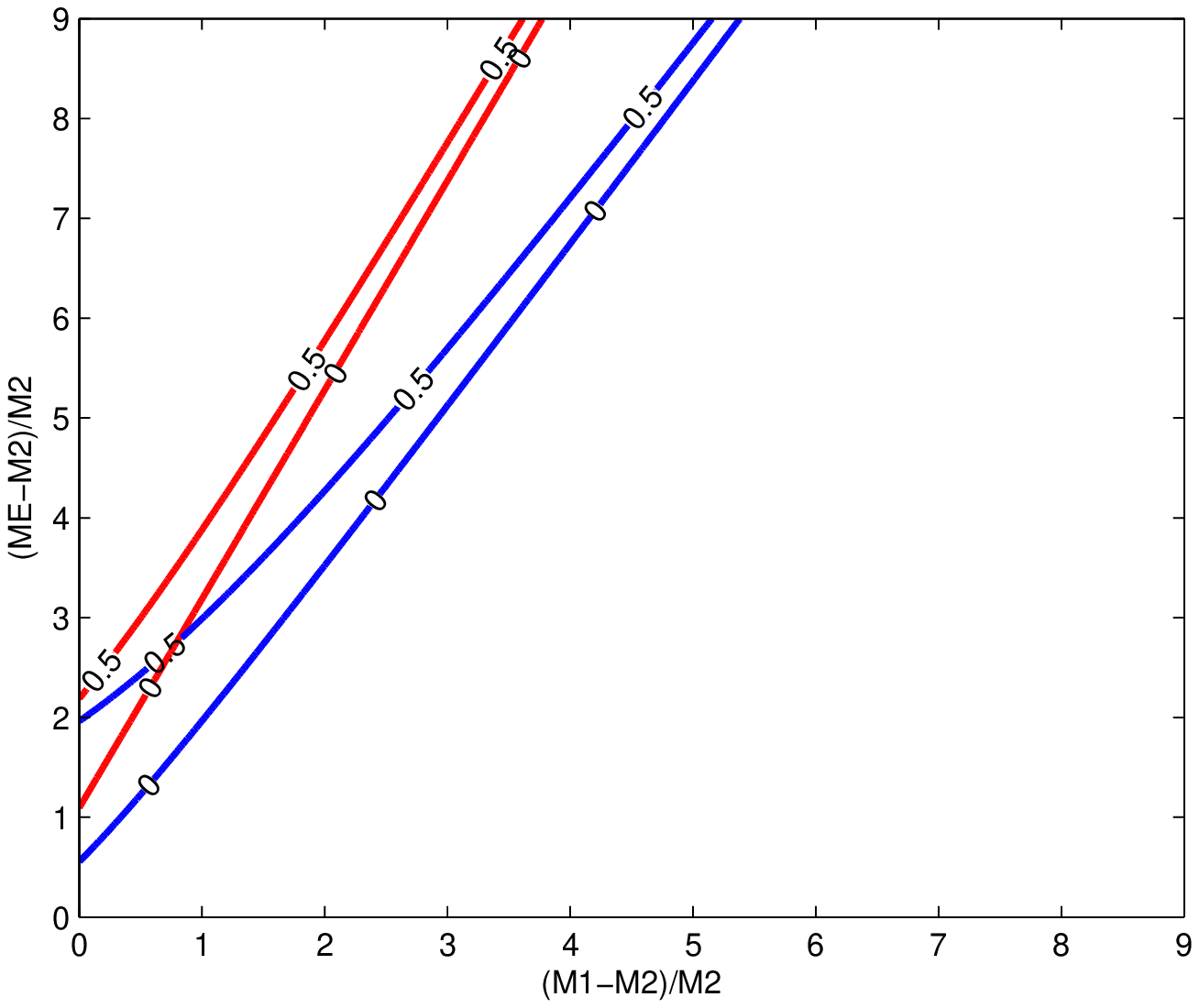}
\caption{Constant $T$ -contours in the $((M_1-M_2),(M_E-M_2)$-plane for the choice of masses $M_2=0.5 M_Z$, (left) and $M_2=M_Z$ (right). Within each panel, the two sets of curves correspond to $\sin\theta=0.1$ (left set) and $\sin\theta=0.5$ (right set).}
\label{Tcontours_rhoneg}
\end{figure}

Then, consider the information on $S$ together with the values of $T$ shown in left and right panels of Fig. \ref{Tcontours_rhoneg}. Within each panel, the two sets of curves correspond to two different values of the mixing angle $\sin\theta$. The relation between the masses $M_1$ and $M_E$ is roughly $M_E\sim 2M_1-M_2$, with the constant of proportionality changing from $2$ to $1.6$ as the value of the mixing angle increases from $\sin\theta=0.1$ to $\sin\theta=0.5$. The results for larger values of $M_2$ fall almost on these same curves and in particular for $M_2<M_1,M_E$ and considering $M_2$ up to 10$M_Z$ the above mentioned relation $M_E\sim 2M_1-M_2$ remains valid.

We observe that $S$-parameter is only modestly restrictive over the mass ranges considered here in comparison to $T$ which provides more stringent constraints. Also note how relatively large mass differences can be accommodated within this model, namely from the above figures one infers that the ratio between the charged lepton and the neutrino masses can easily be a factor of ten, while the values of $S$ and $T$ are, respectively, 0.1 and 0.5 which are consistent with current data. 

\begin{figure}[htb]
\centerline{\includegraphics[width=10cm]{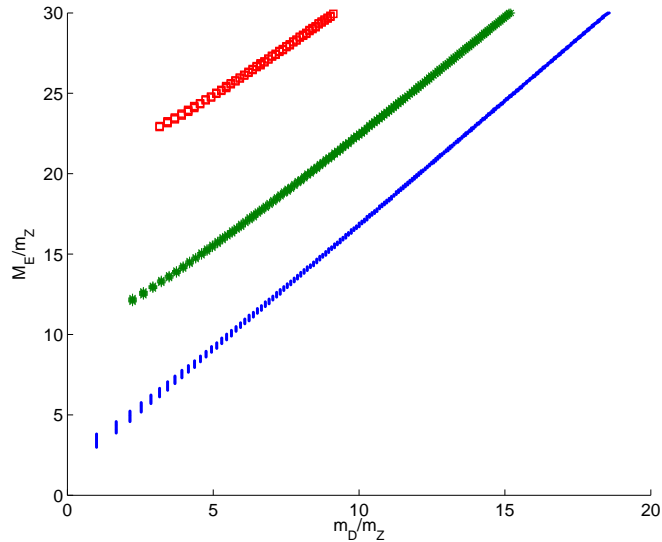}}
\caption{Values of $M_E$ and $m_D$ allowed by constraints $|S|\le 0.2$ and $0<T<0.5$ analysis. On different bands of points $M_L$ and $M_R$ take different fixed values. From bottom to top: $-M_L=M_R=M_Z$ (dots), $-M_L=M_R/5=M_Z$ (stars) and $-M_L=M_R/10=M_Z$ (squares).} 
\label{mass_constraints}
\end{figure}

Overall, from this section we conclude that MWTC model with most general mass spectrum for the fourth generation leptons is compatible with the current precision data on the electroweak observables. Furthermore, the approximate estimate $M_E\sim 2M_N$ between the masses of the charged lepton and the heavier neutrino remains as a good guiding rule of thumb within the spectrum similarly to the already established special cases of Dirac \cite{Dietrich:2005jn} and purely left handed Majorana \cite{Kainulainen:2006wq} neutrinos. 
Our analysis implies that the precision observables do not impose a strong preference towards particular neutrino mass pattern. Namely, for any values of $M_L$ and $M_R$ by adjusting $m_D$ and $M_E$ accordingly one can find portions of parameter space where $S$ and $T$ will satisfy the experimental constraints and all mass eigenstates, $E$, $\chi_1$ and $\chi_2$, are heavy enough to have escaped direct detection so far. As an example, in Fig. \ref{mass_constraints} we show the values of $M_E$ and $m_D$ allowed by restricting $|S|<0.3$ and $0<T<0.5$ with different sets of points corresponding to different hierarchy between $M_L$ and $M_R$. The lowermost band corresponds to $-M_L=M_R=M_Z$, the middle band corresponds to $-M_L=M_R/5=M_z$ and the upper band corresponds to $-M_L=M_R/10=M_Z$. From the figure it would seem at first that as the hierarchy between $M_L$ and $M_R$ increases, the allowed values of $M_E$ and $m_D$ also increase but this actually follows from two effects: First, the values of $M_E$ and $m_D$ reflect the overall value of $M_L$ and $M_R$ and not their ratio. For example, if we set $M_L=0.1 M_Z$ and $M_R=5 M_z$ the resulting $M_E(m_D)$ curve would lie on top of the middle curve in Fig. \ref{mass_constraints} although the corresponding ratio $M_L/M_R$ differs by order of magnitude between the two cases. Second the results are also affected by the requirement that $M_E>M_1>M_2$ for the cases which we consider in this paper. 

In general, as noted already in the beginning of this section, the possible mass splittings within the fourth generation leptons are important in achieving agreement with precision data. For suitable values of the masses the leptons will generate negative contribution to $S$ which helps to partly compensate for positive contribution due to techniquarks. Generally the number of techniquark doublets is large and a single doublet of leptons does typically not provide a big enough contribution to cancel big enough portion of the techniquark contribution. In the MWTC model considered here it is also important in this respect that there are only three technidoublets contributing to the $S$-parameter which is therefore small to begin with. The split masses of the leptons will also lead to nonzero contribution to the $T$-parameter. Together with positive overall contribution to $S$ it is important to have also overall positive contribution to $T$ since the data showns preference for the $T\sim S$ direction.

However, we note the well known fact for Majorana neutrinos, that $T$-parameter can be negative over a significant portion of the parameter space in contrast to the Dirac case where the $T$-parameter is manifestly positive definite. While this feature is not of direct interest for the model study here, it might be useful in models with extended particle content which would also yield further positive contributions to the $T$-parameter.  

In the following sections we will consider the phenomenological implications of the fourth generation neutrinos in the MWTC model. As a starting point for their searches at the LHC we shall take the masses of the charged lepton and the heavier neutrino eigenstate to be in the range of the electroweak scale and consider the lightest neutrino state to have a mass of $M_Z/2$ or $M_Z$. These cases provide a natural starting point for the searches at LHC although consistency with precision data also allows for relatively heavy leptons, even up to the TeV range.

\subsection{Production of new leptons}

Given the analysis of the previous section, the favored range for the masses of the fourth generation leptons is of the order of the electroweak scale. Therefore these leptons serve as an important probe of this model at the LHC, as discussed in e.g. \cite{Holdom:2006mr}. A Simple process which comes to mind is pair production of charged fourth generation leptons. However, the production of the fourth generation neutrinos, $\bar{\chi}\chi$, may be more interesting than production of $\bar{E}E$ pair. This is so since on the one hand the neutrino is expected to be lighter than the charged lepton and on the other hand its decay modes may provide more interesting observables in case that the fourth generation neutrino is not absolutely stable. For example, consider production of a pair of neutrinos, and subsequent decay $\chi\rightarrow \ell W$, where $\ell=\mu,\tau$ or $e$, then the possible final states are
\begin{itemize}
\item $2\ell+4{\textrm{jets}}$
\item $3\ell+2{\textrm{jets}}+\Eslash$
\item $4\ell+\Eslash$.
\end{itemize}
Apparently the third one is similar to what one would expect from the decay of a pair of neutralinos, but the other two should provide a way to distinguish neutrinos from neutralinos. The first one appears interesting since two same-sign leptons can appear in the final state due to the fact that the initial neutrinos are Majorana particles. In addition to $\chi\chi$ production, $2\ell$ and $3\ell$ signals may also be generated in charged current $\ell^{\pm}\chi$ production channel. Bearing this in mind, in the following we focus on the $\ell^{\pm}\ell^{\pm}$ and $\ell^{\pm}\ell^{\pm}\ell^{\mp}$ final states arising from $Z^* \to \chi\chi$ and $W^{*} \to \ell^{\pm}\chi$ production channels.

We assume that the SM neutrinos are Majorana particles which will only affect the partial width of the heavy neutrinos in the $\chi\to Z\nu$ decay. Decay channel to Higgs is assumed to be kinematically forbidden for the neutrino masses we consider in this section; see also Fig.\ref{Higgsfrac}. Feynman rules are given in the appendix of \cite{delAguila:2008cj} in Table 31. We take  off-diagonal $V_{l\chi}$ lepton mixing angle to be real and consider $\rho_1=\rho_2=1$ case, for simplicity. In accordance with the notation of previous sections, we will generically call our two heavy Majorana neutrinos as $\chi_1$ and $\chi_2$ (with $\chi_1$ being the heavy state).

Following \cite{delAguila:2008cj}, we assume that $\chi_1$-neutrino couples to muons only and saturates the latest experimental bound on off-diagonal lepton mixing element $|V_{\mu \chi}|^2<0.0032$ \footnote{Note that this experimentally constrained factor contains contribution from the flavor mixing between generations and also the contribution $\sim\cos\theta$ from the mixing between left- and right handed fourth generation neutrino states.}. In our specific examples, $\chi_2$-neutrino will decouple by either being dominantly right-handed state or by assuming it being lighter than the W and Z and, thus, decaying via off-shell gauge bosons. 

\begin{figure}[htb]
\includegraphics[width=0.5\textwidth]{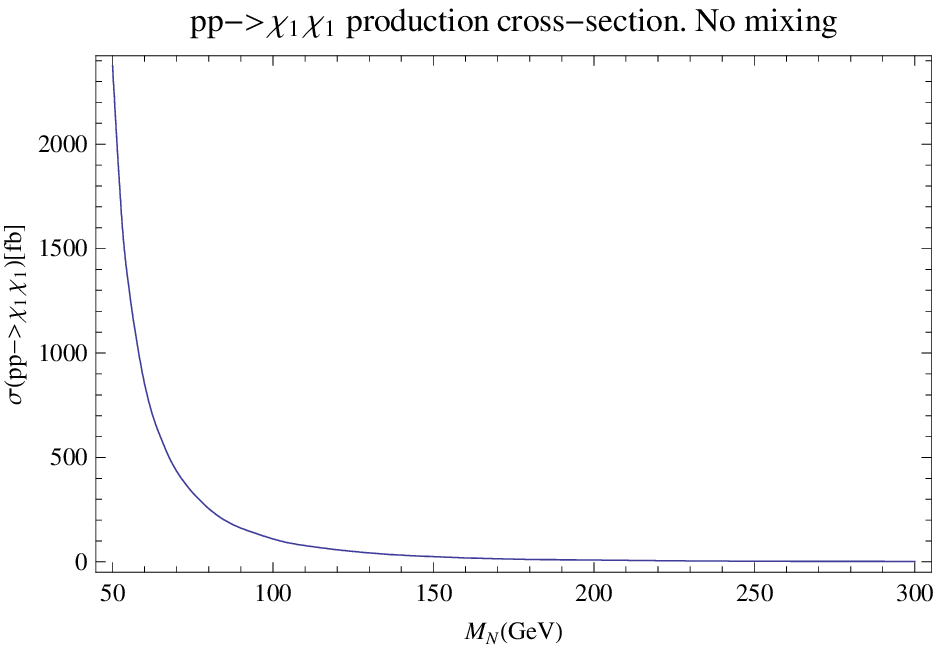}
\includegraphics[width=0.5\textwidth]{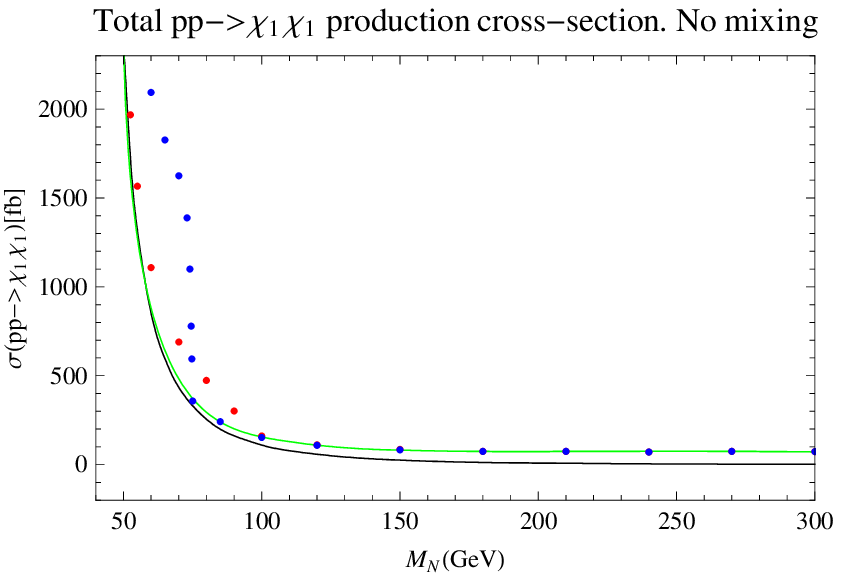}
\caption{ (left) 2 $\to$ 2 cross section for heavy neutrino production $q\bar{q} \to \chi_1\chi_1$ for sin$\theta=0$ via off-shell Z boson channel; (right) total heavy neutrino production $q\bar{q} \to \chi_1\chi_1$ for sin$\theta=0$ including vector boson fusion mechanism. Higgs masses are 100 GeV (solid green), 150 GeV (blue-dotted), 200 GeV (red-dotted). Solid black is the same as on the left figure.}
\label{prod}
\end{figure}

In the left panel of Fig.\ref{prod} we plot the $\sigma (pp \to \chi_1\chi_1)$ production cross-section as a function of $\chi_1$-neutrino mass for sin$\theta$=0. For a typical type I seesaw scenario this cross-section is very small because it requires to mix SM neutrinos with heavy neutrinos twice which suppresses cross-section by a $|V_{\ell \chi}|^4$. Corresponding $\sigma (pp \to \ell^{\pm}\chi_1)$ production cross-section can be found, for example, in \cite{delAguila:2008cj} and we confirmed it in our numerical simulation. For every scenario considered later in this section we show contributions from $\ell^{\pm}\chi_1$ and $\chi_1\chi_1$ production channels separately. 

Our first scenario deals with same-sign leptons in the final state and Table \ref{numerics} illustrates this scenario numerically. First two specific realizations of this scenario feature $\chi_2$-neutrino being the right-handed state (sin$\theta=0$) case with $\chi_1$-neutrino mass equal to 90 GeV and 100 GeV. These masses were chosen to go in parallel with S and T analysis of the previous section ($M_V\equiv M_2=M_Z/2$) and for easy comparison with \cite{delAguila:2008cj} (100 GeV mass case). Third realization with $M_N$=135 GeV and mixing angle sin$\theta=0.5$ was selected to probe heavier neutrino masses. Again, in this last realization, the $\chi_2$-neutrino is assumed to decay via off-shell gauge bosons and, as such, being neglected in the analysis. 

For event pre-selection we require the presence of two like-sign charged leptons with transverse momentum larger than 30 GeV, and an additional lepton of opposite charge. The choice of the $p_T$ cut for like-sign leptons is motivated by the need to reduce backgrounds where soft leptons are produced in $b$ decays, for example $t \bar t nj$ ($nj$ standing for $n$ jets) in the dilepton channel. For the final event selection we also require:
\begin{itemize}
\item[(i)] at least two jets in the final state with $p_T > 20$ GeV, and no $b$-tagged jets;
\item[(ii)] missing energy smaller than 30 GeV;
\item[(iii)] the transverse angle between the two leptons must be larger than $\pi/2$.
\end{itemize}

The signal and SM background cross sections for these two stages of event selection are also given in Table~\ref{numerics}. The number of the selected events for an integrated luminosity of 10 fb$^{-1}$ and the corresponding statistical significance are also presented there. We observe that 100 GeV and 135 GeV neutrinos may be discovered early under these conditions.

\begin{table}
\caption{Signal cross-sections $\sigma$(in fb) 
with the corresponding leading SM background for three scenarios described in the text. 
Pre-selection and selection criteria are also described in the text.}
\label{numerics}
\begin{center}
\begin{tabular}{|c|c|c|c|c|}
\hline
$\mu^{\pm}\mu^{\pm}$ X ($M_{\chi_1}$=100 GeV)&Pre-selected $\sigma$(fb)&Selected $\sigma$(fb)&Events/10 fb$^{-1}$&S/$\sqrt{B}$\\
\hline
$\mu^{\pm}\chi_1:\mu^{\pm}\mu^{\pm}$ + 2 jets  &1.11&0.6&6&7.25\\
\hline
$\chi_1\chi_1:\mu^{\pm}\mu^{\pm}$ + 4 jets  &2.5&1.32&13.2&\\
\hline
SM background &10.25&0.7&7&\\
\hline

\hline
$\mu^{\pm}\mu^{\pm}$ X ($M_{\chi_1}$=90 GeV)&Pre-selected $\sigma$(fb)&Selected $\sigma$(fb)&Events/10 fb$^{-1}$&S/$\sqrt{B}$\\
\hline
$\mu^{\pm}\chi_1:\mu^{\pm}\mu^{\pm}$ + 2 jets  &0.21&0.113&1.1&1.89\\
\hline
$\chi_1\chi_1:\mu^{\pm}\mu^{\pm}$ + 4 jets  &0.73&0.39&3.9&\\
\hline
SM background &10.25&0.7&7&\\
\hline

\hline
$\mu^{\pm}\mu^{\pm}$ X ($M_{\chi_1}$=135 GeV)&Pre-selected $\sigma$(fb)&Selected $\sigma$(fb)&Events/10 fb$^{-1}$&S/$\sqrt{B}$\\
\hline
$\mu^{\pm}\chi_1:\mu^{\pm}\mu^{\pm}$ + 2 jets  &2.1&1.1&11&6.95\\
\hline
$\chi_1\chi_1:\mu^{\pm}\mu^{\pm}$ + 4 jets  &1.4&0.74&7.4&\\
\hline
SM background &10.25&0.7&7&\\
\hline
\hline
$\ell^{\pm}\ell^{\pm}\ell^{\mp}$ X ($M_{\chi_1}$=100 GeV)&Pre-selected $\sigma$(fb)&Selected $\sigma$(fb)&Events/10 fb$^{-1}$&S/$\sqrt{B}$\\
\hline
$\mu^{\pm}\chi_1:\ell^{\pm}\ell^{\pm}\ell^{\mp}$ + $\Eslash$  &1.95&1.52&15.2&12.15\\
\hline
$\chi_1\chi_1:\ell^{\pm}\ell^{\pm}\ell^{\mp} $+ 2 jets+ $\Eslash$  &4.10&3.20&32&\\
\hline
SM background &76.7&1.51&15.1&\\
\hline
\end{tabular}
\end{center}
\end{table}

In the second scenario we consider trilepton final state with $\ell=e,\mu$, $M_{\chi_1}$=100 GeV and all other conditions as in the first scenario. 

Trilepton signals can be produced in the two charged current decay channels of the heavy neutrino, with subsequent leptonic decay of the $W$ boson, {\em e.g.}
\begin{align}
& \ell^+ \chi_1 \to \ell^+ \ell^- W^+ \to \ell^+ \ell^- \ell^+ \bar \nu \,, \nonumber \\
& \ell^+ \chi_1 \to \ell^+ \ell^+ W^- \to \ell^+ \ell^+ \ell^- \nu \,.
\end{align}

They can also be produced in the $\chi_1\chi_1$ production channel {\em e.g.}
\begin{align}
&  \chi_1\chi_1 \to \ell^+ \ell^- W^+  W^- \to \ell^+ \ell^- \ell^- \bar \nu + 2 \text{ jets} \,, \nonumber \\
&  \chi_1\chi_1 \to \ell^+ \ell^+ W^-  W^- \to \ell^+ \ell^+ \ell^- \bar \nu + 2 \text{ jets} \,,
\end{align}
(and small additional contributions from $\tau$ leptonic decays for both production channels).

This trilepton final state is very clean once that $WZnj$ production can be almost eliminated with a simple cut on the invariant mass of opposite charge leptons.

For event pre-selection we again require two same-sign charged leptons with $p_T> 30 $ GeV.
For event selection, we require that neither of the two opposite-sign lepton pairs have an invariant mass closer to $M_Z$ than 10 GeV and we ask that
\begin{itemize}
\item[(i)] no $b$ jets can be present in the final state;
\item[(ii)] the like-sign leptons must be back-to-back, with their angle in transverse plane larger than $\pi/2$.
\end{itemize}
Our results are shown in Table \ref{numerics} and again we notice that trilepton channel may cross-check potential discovery scenario in dilepton channel.

Up to this point we only considered pp$\to \chi_1\chi_1(\chi_2\chi_2)$ neutrino production via off-shell Z boson. In the right panel of Fig.\ref{prod} we include the vector boson fusion channel to the Higgs or Z boson under the same conditions as in left panel of the same figure. We observe the enhancement in the low neutrino mass region due to on-shell Higgs decay to the pair of $\chi_1$-neutrinos. Higher neutrino mass region is also enhanced due to the vector boson fusion to the Z boson which also consequently decay to the pair of $\chi_1$-neutrinos. If the Higgs mass happens to be in the specified regions, this additional production channel would modify the corresponding numbers in Table \ref{numerics} with an appropriate multiplicative factor. Similar enhancement might also occur in the charged current production channel pp $\to \ell^{\pm}\chi_1$.

However, for the masses considered in the Table \ref{numerics}, significant enhancements would occur only for parameters values where Higgs is heavy enough to produce two neutrinos and, at the same time, neutrinos are heavy enough to decay to the on-shell W and Z. For example, 150 GeV Higgs case considered in the right panel of Fig \ref{prod} would not satisfy this condition, and only 200 GeV Higgs mass case would give an enhancement. Thus, for example, 90 GeV $\chi_1$-neutrino in the first scenario would, approximately, receive an additional multiplicative factor of 2 in all the corresponding numbers in Table \ref{numerics}.

\subsection{Higgs decay}

Since the effective coupling between the composite scalar sector and standard model matter fields in the MWTC theory is simple, let us also outline possible effects that the new leptons would have on the decays of the composite Higgs boson. Since the coupling between the composite Higgs and matter fields is only effective one, these results should be taken as qualitative ones illustrating possible effects which can be expected to arise. There are several different possibilities depending on the masses of the new leptons and the composite Higgs. Since the new charged lepton is constrained to have mass $M_E\ge 2M_z$, its effect looks similar to that of the top quark. Most interesting implications are due to the new neutrino which can be relatively light, but couple only weakly to electroweak currents and hence evade the LEP bounds. From equations (\ref{gauge_interactions}) and (\ref{higgs_interactions}) it follows that depending on the neutrino and higgs masses, the decay rates can be dramatically affected by the existence of such fourth generation neutrino. The dominant effect for the Higgs decay is insensitive to the magnitude of weak interactions, i.e. the neutrino mixing angle, since the Higgs field couples with the strength proportional to the mass of these particles. However, as can be seen from (\ref{higgs_interactions}), there is also a contribution depending on the mixing angle. Also the sign of $\rho_1\rho_2$ affects the Higgs couplings, but qualitatively the effects are similar both for $\rho_1\rho_2=1$ and for $\rho_1\rho_2=-1$.

\begin{figure}[htb]
\includegraphics[width=0.5\textwidth]{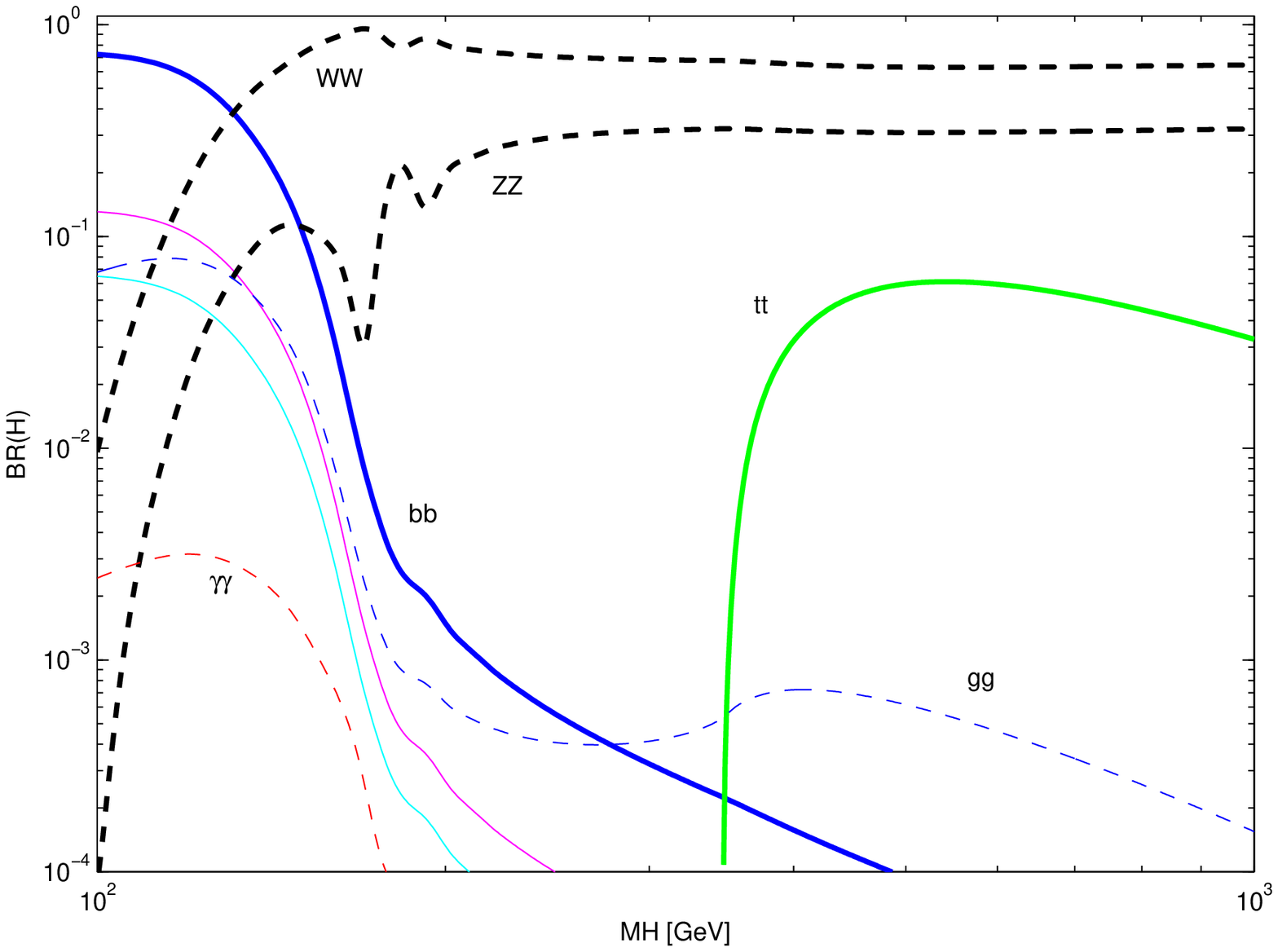}
\includegraphics[width=0.5\textwidth]{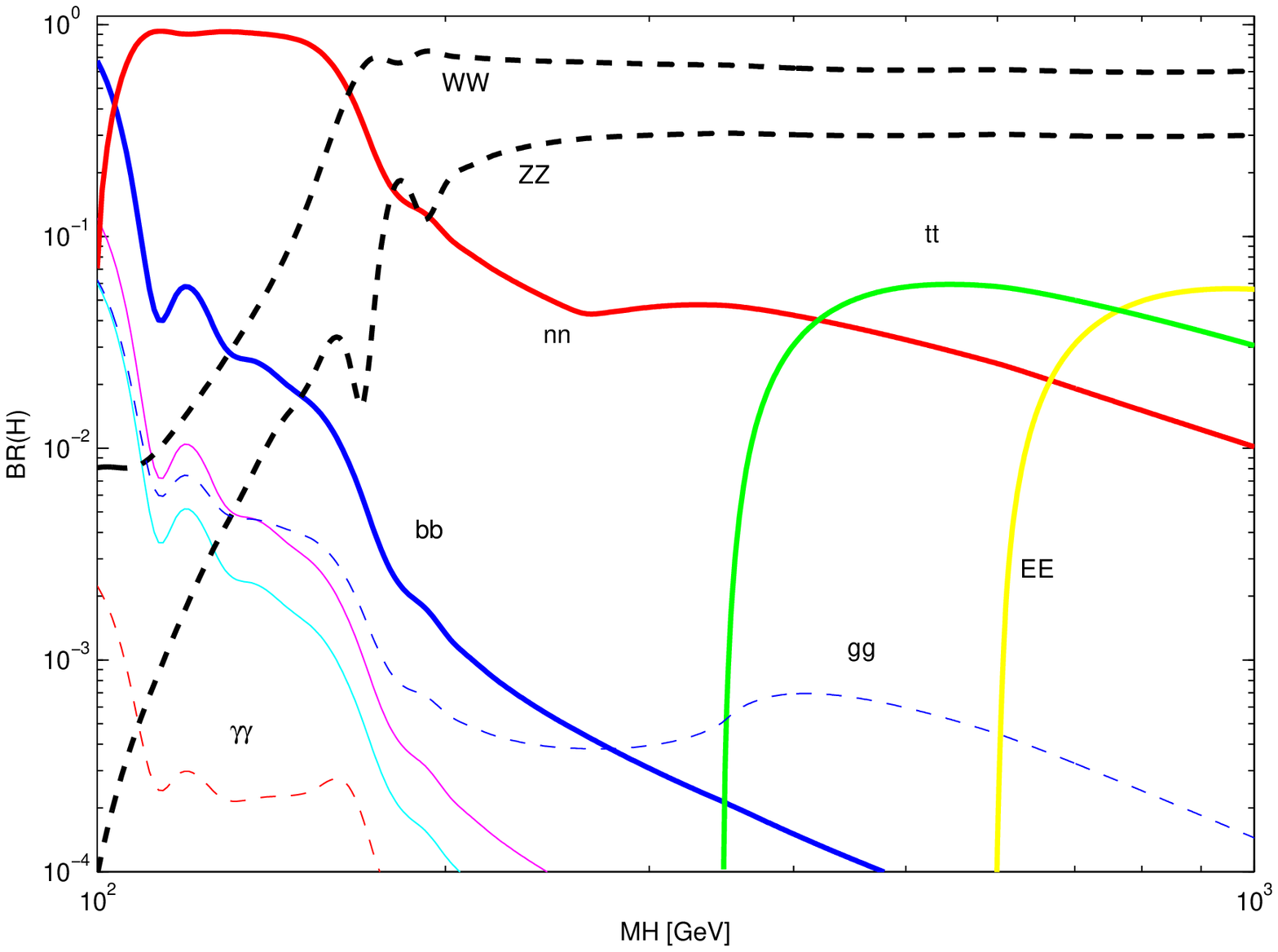}
\caption{The branching ratios for the decay of the Higgs boson in Standard Model (left panel) and in MWTC (right panel)}
\label{Higgsfrac}
\end{figure}

In Fig. \ref{Higgsfrac} we show the branching ratios of the Higgs boson for different final states as a function of the Higgs boson mass. The left panel shows the familiar figure corresponding to the final states present in the standard model. In the right panel on the other hand we have taken into account the new leptons in the MWTC model. If the lighter neutrino state in the fourth generation is around $M_Z/2$, it will create an important channel significantly reducing the contribution of the other final states relevant for the light Higgs searches at the LHC. 

To compute the contribution in MWTC, we set $M_E=300$ GeV, $M_1=130$ GeV and $M_2=50$ GeV, $\sin\theta=0.45$ and $\rho_2=+1$ which gives acceptable values $(S,T)=(0.17,0.6)$. It is clear that realization of this particular scenario requires a significant amount of fine tuning regarding the masses of the fourth generation leptons as well as the mass of the Higgs boson, and we only point this out as an interesting special case. Furthermore, there are likely to be important corrections to Higgs decay due to the underlying strong dynamics. However, assuming that the composite SM-like Higgs is the lightest technihadron and for neutrino masses below $M_Z$ our effective vertices should provide qualitatively correct picture. Careful studies of various effects relevant for light Higgs are important since there is some theoretical and phenomenological bias towards this mass range; also due to recent experimental results from the Tevatron \cite{Phenomena:2009pt}. 

Also the Higgs production is modified from that of the standard model. Here it should be stressed that this modification is different for the case here and for the case of sequential fourth generation where also a QCD-like heavy quark generation appears. In the latter case the new QCD-matter doublets enhance the gluon-gluon fusion due to new degrees of freedom available to run in the loop coupling the gluons with the Higgs \cite{Kribs:2007nz}. In the former case relevant here no such QCD contribution exists. Rather, in this case the modifications to the Higgs production rates arise from the resonant weak boson scattering. These different production channels of spin-0 resonance at LHC will provide important tool to discriminate between these two different possibilities if a fourth generation of standard model -like matter appears in nature. In \cite{Antipin:2009ch} similar reasoning has been applied to other theories which lead to appearance of spin-2 resonances.

\section{Conclusions and outlook}
\label{checkout}

In this paper we have analyzed the contributions to the precision parameters from fourth generation leptons and in particular considered phenomenological implications of a novel technicolor scenario where a fourth family of leptons arises without new QCD quarks which would appear in typical sequential extensions of the Standard Model. Allowing for the most general mass structure for these leptons we have evaluated the $S$ and $T$ parameters and shown how these constrain the masses of the new fourth generation leptons. We have discussed some collider signatures which can be used to probe the existence of these new leptons at the LHC. We have shown that the decay rates of the Higgs particle can be affected through the decays to these new particles in the case of light Higgs particle.

We chose to study the minimal walking technicolor model, since it is naturally required to contain a full fourth generation in order to saturate the Witten anomaly. The effects from the leptonic sector are novel since the quark sector does not carry QCD color but is instead technicolored. The technicolor sector will provide its own characteristic signals which are probably best studied through $WW$ scattering, while the new leptons will be manifest through other initial states as well. 

Here we have concentrated on the typical mass hierarchy of the leptons, i.e. charged lepton always heavier than the lightest neutrino. However, one can also entertain a thought that the charged state would be the lightest. Then its decay could only proceed via mixing with the lighter generations, making it relatively long lived and perhaps amendable even to direct observation. Also, we have considered exclusively the standard model like hypercharge assignments for the techniquarks and fourth generation leptons. However, saturation of gauge anomalies in MWTC theory allows for other possibilities which should be studied in detail. For example, the "neutrino" could have charge $-e$ while the other lepton would be doubly charged. 

There are also other theories which one can consider. For example, another promising candidate for strong dynamics is a SU(3) gauge theory with two flavors of sextet techniquarks. Also this particle content should lead to walking behavior of the technicolor coupling constant as was proposed in \cite{Sannino:2004qp}. For a lattice simulations concerning the running of the coupling in this model see \cite{Shamir:2008pb}, and for a study of the associated collider phenomenology see \cite{Belyaev:2008yj}. In this model one does not need to include additional leptons to saturate Witten anomaly and also the naive $S$-parameter is larger than in the minimal model; namely $S\sim 1/\pi$ which is still marginally within the experimental bounds but $T=0$ if degenerate techniquarks are assumed. However, if one insists on standard model like hypercharge assignments for the techniquarks, then, to saturate the gauge anomalies, the introduction of two lepton doublets would be necessary. From the weak interaction viewpoint one could regard the model as one with two new generations. Now, given this extended particle content the $S$ and $T$ can be made better compatible with the existing data since with appropriate masses for the new two lepton generations generate negative contribution to $S$ making it smaller than $1/\pi$ and positive contribution to $T$ similarly to what happens in the minimal model which we have considered here. The collider signatures of the (lightest) additional leptons are likely to be similar as the ones considered here. 

Hence, the existence of fourth family of matter analogous to the three currently featured in the standard model remains as a simple and theoretically motivated possibility of new physics appearing at the LHC. If indeed such matter content appears, the story is likely to be richer than just a sequential fourth generation. 

\acknowledgments
We thank K.~Kainulainen and J.~Virkaj\"arvi for discussions. M.H. was supported by a grant from M.~Ehrnrooth foundation.


\begin{thebibliography}{99}

\bibitem{Lunghi:2009sm}
  E.~Lunghi and A.~Soni,
  {\em Hints for the scale of new CP-violating physics from B-CP anomalies},
  arXiv:0903.5059 [hep-ph].

\bibitem{Soni:2008bc}
  A.~Soni, A.~K.~Alok, A.~Giri, R.~Mohanta and S.~Nandi,
  {\em The Fourth family: A Natural explanation for the observed pattern of
  anomalies in $B^-$ CP asymmetries},
  arXiv:0807.1971 [hep-ph].

\bibitem{Frampton:1999xi}
  P.~H.~Frampton, P.~Q.~Hung and M.~Sher,
  {\em Quarks and leptons beyond the third generation},
  Phys.\ Rept.\  {\bf 330}, 263 (2000)
  [arXiv:hep-ph/9903387].

\bibitem{Holdom:2006mr}
  B.~Holdom,
  {\em The discovery of the fourth family at the LHC: What if?},
  JHEP {\bf 0608}, 076 (2006)
  [arXiv:hep-ph/0606146].

\bibitem{Bobrowski:2009ng}
   M.~Bobrowski, A.~Lenz, J.~Riedl and J.~Rohrwild,
   arXiv:0902.4883 [hep-ph].

\bibitem{TC}
S.~Weinberg,
{\em Implications Of Dynamical Symmetry Breaking: An Addendum},
Phys.\ Rev.\ D {\bf 19}, 1277 (1979);
L.~Susskind,
{\em Dynamics Of Spontaneous Symmetry Breaking In The Weinberg-Salam Theory},
Phys.\ Rev.\ D {\bf 20}, 2619 (1979);
  E.~Eichten and K.~D.~Lane,
  {\em{Dynamical Breaking Of Weak Interaction Symmetries}},
  Phys.\ Lett.\  B {\bf 90}, 125 (1980).

\bibitem{Hill:2002ap}
C.~T.~Hill and E.~H.~Simmons,
{\em Strong dynamics and electroweak symmetry breaking},
Phys.\ Rept.\  {\bf 381}, 235 (2003) [Erratum-ibid.\  {\bf 390},
553 (2004)];
F.~Sannino,
{\em Dynamical Stabilization of the Fermi Scale: Phase Diagram of Strongly
Coupled Theories for (Minimal) Walking Technicolor and Unparticles},
arXiv:0804.0182 [hep-ph].

\bibitem{Holdom:1981rm}
B.~Holdom,
{\em{Raising The Sideways Scale}},
Phys.\ Rev.\ D {\bf 24}, 1441 (1981).

\bibitem{Yamawaki:1985zg}
K.~Yamawaki, M.~Bando and K.~i.~Matumoto,
{\em{Scale Invariant Technicolor Model And A Technidilaton}},
Phys.\ Rev.\ Lett.\  {\bf 56}, 1335 (1986).

\bibitem{Appelquist:an}
T.~W.~Appelquist, D.~Karabali and L.~C.~R.~Wijewardhana,
{\em{Chiral Hierarchies And The Flavor Changing Neutral Current Problem In
Technicolor}},
Phys.\ Rev.\ Lett.\  {\bf 57}, 957 (1986);
T.~Appelquist, A.~Ratnaweera, J.~Terning and
L.~C.~R.~Wijewardhana,
{\em{The phase structure of an SU(N) gauge theory with N(f) flavors}},
Phys.\ Rev.\ D {\bf 58}, 105017 (1998).

\bibitem{Sannino:2004qp}
  F.~Sannino and K.~Tuominen,
  {\em Orientifold theory dynamics and symmetry breaking},
  Phys.\ Rev.\  D {\bf 71}, 051901 (2005)
  [arXiv:hep-ph/0405209].

\bibitem{Dietrich:2005jn}
  D.~D.~Dietrich, F.~Sannino and K.~Tuominen,
  {\em Light composite Higgs from higher representations versus electroweak
  precision measurements: Predictions for LHC},
  Phys.\ Rev.\  D {\bf 72}, 055001 (2005)
  [arXiv:hep-ph/0505059];
  D.~D.~Dietrich, F.~Sannino and K.~Tuominen,
  {\em Light composite Higgs and precision electroweak measurements on the Z
  resonance: An update},
  Phys.\ Rev.\  D {\bf 73}, 037701 (2006)
  [arXiv:hep-ph/0510217].


\bibitem{Catterall:2007yx}
  S.~Catterall and F.~Sannino,
  {\em{Minimal walking on the lattice}},
  Phys.\ Rev.\  D {\bf 76}, 034504 (2007)
  [arXiv:0705.1664 [hep-lat]];
  L.~Del Debbio, A.~Patella and C.~Pica,
  {\em{Higher representations on the lattice: numerical simulations. SU(2) with
  adjoint fermions}},
  arXiv:0805.2058 [hep-lat];
  
\bibitem{Catterall:2008qk}
  S.~Catterall, J.~Giedt, F.~Sannino and J.~Schneible,
  {\em{Phase diagram of SU(2) with 2 flavors of dynamical adjoint quarks}},
  arXiv:0807.0792 [hep-lat].

\bibitem{Hietanen:2008mr}
  A.~J.~Hietanen, J.~Rantaharju, K.~Rummukainen and K.~Tuominen,
  {\em Spectrum of SU(2) lattice gauge theory with two adjoint Dirac flavours},
  arXiv:0812.1467 [hep-lat].

\bibitem{Hietanen:2009az}
  A.~J.~Hietanen, K.~Rummukainen and K.~Tuominen,
  {\em Evolution of the coupling constant in SU(2) lattice gauge theory with two
  adjoint fermions},
  arXiv:0904.0864 [hep-lat].

\bibitem{Witten:1982fp}
  E.~Witten,
  {\em An SU(2) anomaly},
  Phys.\ Lett.\  B {\bf 117}, 324 (1982).

\bibitem{Akerib:2004fq}
  D.~S.~Akerib {\it et al.}  [CDMS Collaboration],
  {\em First results from the cryogenic dark matter search in the Soudan
  Underground Lab},
  Phys.\ Rev.\ Lett.\  {\bf 93}, 211301 (2004)
  [arXiv:astro-ph/0405033].

\bibitem{Kainulainen:2006wq}
  K.~Kainulainen, K.~Tuominen and J.~Virkajarvi,
  {\em The WIMP of a minimal technicolor theory},
  Phys.\ Rev.\  D {\bf 75}, 085003 (2007)
  [arXiv:hep-ph/0612247].

\bibitem{Angle:2008we}
  J.~Angle {\it et al.},
  {\em Limits on spin-dependent WIMP-nucleon cross-sections from the XENON10
  experiment},
  Phys.\ Rev.\ Lett.\  {\bf 101}, 091301 (2008)
  [arXiv:0805.2939 [astro-ph]].

\bibitem{Kouvaris:2007iq}
  C.~Kouvaris,
  {\em Dark Majorana Particles from the Minimal Walking Technicolor},
  Phys.\ Rev.\  D {\bf 76}, 015011 (2007)
  [arXiv:hep-ph/0703266].

\bibitem{KTVII}
	K.~Kainulainen, K.~Tuominen and J.~Virkaj\"arvi, in progress.

\bibitem{Foadi:2007ue}
  R.~Foadi, M.~T.~Frandsen, T.~A.~Ryttov and F.~Sannino,
  {\em Minimal Walking Technicolor: Set Up for Collider Physics},
  Phys.\ Rev.\  D {\bf 76}, 055005 (2007)
  [arXiv:0706.1696 [hep-ph]];
  R.~Foadi, M.~T.~Frandsen and F.~Sannino,
  {\em Constraining Walking and Custodial Technicolor},
  Phys.\ Rev.\  D {\bf 77}, 097702 (2008)
  [arXiv:0712.1948 [hep-ph]].

\bibitem{Barroso:1983rd}
  A.~Barroso and J.~Maalampi,
  Phys.\ Lett.\  B {\bf 132}, 355 (1983).

\bibitem{He:2001tp}
  H.~J.~He, N.~Polonsky and S.~f.~Su,
  {\em Extra families, Higgs spectrum and oblique corrections},
  Phys.\ Rev.\  D {\bf 64}, 053004 (2001)
  [arXiv:hep-ph/0102144].

\bibitem{Bertolini:1990ek}
  S.~Bertolini and A.~Sirlin,
  {\em Effect Of A Fourth Fermion Generation On The M(T) Upper Bound},
  Phys.\ Lett.\  B {\bf 257}, 179 (1991).

\bibitem{Gates:1991uu}
  E.~Gates and J.~Terning,
  {Negative Contributions To S From Majorana Particles},
  Phys.\ Rev.\ Lett.\  {\bf 67}, 1840 (1991).

\bibitem{Holdom:1996bn}
  B.~Holdom,
  {\em Negative T?},
  Phys.\ Rev.\  D {\bf 54}, 721 (1996)
  [arXiv:hep-ph/9602248].
  
\bibitem{Kniehl:1992ez}
  B.~A.~Kniehl and H.~G.~Kohrs,
  {\em Oblique radiative corrections from Majorana neutrinos},
  Phys.\ Rev.\  D {\bf 48}, 225 (1993).

\bibitem{Kribs:2007nz}
  G.~D.~Kribs, T.~Plehn, M.~Spannowsky and T.~M.~P.~Tait,
  {\em Four generations and Higgs physics},
  Phys.\ Rev.\  D {\bf 76}, 075016 (2007)
  [arXiv:0706.3718 [hep-ph]].

\bibitem{Ryttov:2008xe}
  T.~A.~Ryttov and F.~Sannino,
  {\em Ultra Minimal Technicolor and its Dark Matter TIMP},
  Phys.\ Rev.\  D {\bf 78}, 115010 (2008)
  [arXiv:0809.0713 [hep-ph]].

\bibitem{Barbieri:2004qk}
  R.~Barbieri, A.~Pomarol, R.~Rattazzi and A.~Strumia,
  {\em Electroweak symmetry breaking after LEP-1 and LEP-2},
  Nucl.\ Phys.\  B {\bf 703}, 127 (2004)
  [arXiv:hep-ph/0405040];
  I.~Maksymyk, C.~P.~Burgess and D.~London,
  {\em Beyond S, T and U},
  Phys.\ Rev.\  D {\bf 50}, 529 (1994)
  [arXiv:hep-ph/9306267].

\bibitem{:2005ema}
    [ALEPH Collaboration and DELPHI Collaboration and L3 Collaboration],
  {\em Precision electroweak measurements on the $Z$ resonance},
  Phys.\ Rept.\  {\bf 427}, 257 (2006)
  [arXiv:hep-ex/0509008].

\bibitem{Amsler:2008zzb}
  C.~Amsler {\it et al.}  [Particle Data Group],
  {\em Review of particle physics},
  Phys.\ Lett.\  B {\bf 667}, 1 (2008).

\bibitem{Peskin:2001rw}
  M.~E.~Peskin and J.~D.~Wells,
  {\em How can a heavy Higgs boson be consistent with the precision  electroweak
  measurements?},
  Phys.\ Rev.\  D {\bf 64}, 093003 (2001)
  [arXiv:hep-ph/0101342].

\bibitem{Denner:1992me}
  A.~Denner, H.~Eck, O.~Hahn and J.~Kublbeck,
  {\em Compact Feynman rules for Majorana fermions},
  Phys.\ Lett.\  B {\bf 291}, 278 (1992).

\bibitem{delAguila:2008cj}
  F.~del Aguila and J.~A.~Aguilar-Saavedra,
  {\em Distinguishing seesaw models at LHC with multi-lepton signals},
  arXiv:0808.2468 [hep-ph].

\bibitem{Phenomena:2009pt}
  T.~N.~Phenomena, H.~w.~group, f.~t.~C.~collaboration and D.~collaboration,
  {\em Combined CDF and DZero Upper Limits on Standard Model Higgs-Boson
  Production with up to 4.2 fb-1 of Data},
  arXiv:0903.4001 [hep-ex].

\bibitem{Shamir:2008pb}
  Y.~Shamir, B.~Svetitsky and T.~DeGrand,
  {\em Zero of the discrete beta function in SU(3) lattice gauge theory with color
  sextet fermions},
  Phys.\ Rev.\  D {\bf 78}, 031502 (2008)
  [arXiv:0803.1707 [hep-lat]].

\bibitem{Antipin:2009ch}
  O.~Antipin and K.~Tuominen,
  {\em Discriminating between technicolor and warped extra dimensional model via
  pp $\to$ ZZ channel},
  arXiv:0901.4243 [hep-ph].

\bibitem{Belyaev:2008yj}
  A.~Belyaev, R.~Foadi, M.~T.~Frandsen, M.~Jarvinen, F.~Sannino and A.~Pukhov,
  {\em Technicolor Walks at the LHC},
  arXiv:0809.0793 [hep-ph].

\end{thebibliography}
\end{document}